\documentclass{iaus}
\usepackage{graphics} 
\title[658 GHz Water Masers]
{658 GHz Vibrationally-Excited Water Masers with the Submillimeter Array}

\author[Hunter et al.]{T.R. Hunter$^1$, K.H. Young$^2$,
R.D. Christensen$^3$, M.A. Gurwell$^2$
} 

\affiliation{$^1$NRAO, 520 Edgemont Rd, Charlottesville, VA 22903, USA \break 
email: thunter@nrao.edu\\ [\affilskip]
$^2$Harvard-Smithsonian Center for Astrophysics, 60 Garden St., Cambridge, MA 02138, USA\\ 
[\affilskip] $^3$Submillimeter Array, 645 North A'ohoku Place,
Hilo, HI 96720, USA}  

\pubyear{2007}
\volume{242}
\pagerange{1--8}
\date{April 2007}
\jname{Proceedings Astrophysical Masers and Their Environments}
\editors{J. Chapman, eds.}
\begin{document}
\maketitle
\begin{abstract}

Discovered in 1995 at the Caltech Submillimeter Observatory (CSO), the
vibrationally-excited water maser line at 658 GHz (455 micron) is seen
in oxygen-rich giant and supergiant stars.  Because this maser can be
so strong (up to thousands of Janskys), it was very helpful during the
commissioning phase of the highest frequency band (620-700 GHz) of the
Submillimeter Array (SMA) interferometer.  From late 2002 to early
2006, brief attempts were made to search for emission from additional
sources beyond the original CSO survey.  These efforts have expanded
the source count from 10 to 16.  The maser emission appears to be
quite compact spatially, as expected from theoretical considerations;
thus these objects can potentially be used as atmospheric phase
calibrators.  Many of these objects also exhibit maser emission in the
vibrationally-excited SiO maser at 215 GHz.  Because both maser
lines likely originate from a similar physical region, these objects
can be used to test techniques of phase transfer calibration between
millimeter and submillimeter bands.  The 658 GHz masers will be 
important beacons
to assess the performance of the Atacama Large Millimeter Array (ALMA)
in this challenging high-frequency band.

\end{abstract}

\firstsection
\section{Introduction}

Water is an asymmetric top molecule with three vibrational quantum
numbers (\cite{Herzberg45}).  The presence of vibrationally-excited
water has been detected at near-IR wavelengths in a variety of
astrophysical environments such as the atmospheres of Mira variable
stars (e.g. R~Leo, \cite{Hinkle79}), semi-regular pulsating stars
(e.g. W~Hya, \cite{Justtanont04}), FU~Orionis objects
(\cite{Reipurth07}), Mars (\cite{Jouglet06}), and in the post-Deep
Impact ejecta from Comet Temple 1 (\cite{Barber06}).  Of the three
vibrational modes, the $v_2=1$ bending mode has the lowest energy
ground state (2297~K). Rotational levels for this mode are tabulated
by \cite{Camy77} and a level diagram is presented by \cite{Alcolea93}.
Frequency measurements are given by \cite{Belov87}, \cite{Pearson91},
and \cite{Chen00}. Analogous frequencies for deuterated water are
currently being measured (\cite{Brunken05}).

\begin{table}
\begin{center}
\caption{Astronomically-detected (sub)millimeter transitions of vibrationally-excited water \label{freq}}
\begin{tabular}{|c|c|c|c|c|c|l|}  
\hline
Frequency  & Rotational & Vibrational &         & $E_{\rm upper}$ & & Discovery\\
(GHz)      & Transition & State       & Species & (K)          & Telescope & Reference\\
\hline
  96.261 & $4_{4,0} \rightarrow 5_{3,3}$ & 010 & para  & 3066 & IRAM 30m & \cite{Menten89}\\
 232.686 & $5_{5,0} \rightarrow 6_{4,3}$ & 010 & ortho & 3465 & IRAM 30m & \cite{Menten89}\\
 293.664 & $6_{6,1} \rightarrow 7_{5,2}$ & 010 & ortho & 2941 & APEX 12m & \cite{Menten06}\\
 336.227 & $5_{2,3} \rightarrow 6_{1,6}$ & 010 & ortho & 2958 & APEX 12m & \cite{Menten06}\\
 658.006 & $1_{1,0} \rightarrow 1_{0,1}$ & 010 & ortho & 2362 & CSO 10.4m & \cite{Menten95}\\
\hline 
\end{tabular}
\end{center}
\end{table}

Emission from water vapor in the $v_2=1$ state has been detected in
astronomical sources in several (sub)millimeter lines (see Table~1).
Searches have been performed for three other $v_2=1$ transitions with
negative results: para $6_{6,0} \rightarrow 7_{5,3}$ at 297.439 GHz
(\cite{Menten06}), ortho 4$_{2,3} \rightarrow 3_{3,0}$ at 12.008 GHz
(\cite{Myers82}), and ortho 4$_{1,4} \rightarrow 3_{2,1}$ at 67.704
GHz (\cite{Pet91}).  All of the lines in Table~\ref{freq} have been
detected in the hypergiant star VY~CMa.  With the exception of the
336~GHz line, all of the detected lines in this object show evidence
for maser action, at least in some velocity components.  The 232 GHz
line is also seen in the semi-regular pulsating star W~Hya.  The
658~GHz transition to the ortho ground state is seen as a strong maser
in both objects, along with eight other evolved stars including three
semi-regular stars (R~Crt, RT~Vir, RX~Boo), two hypergiants (VX~Sgr,
NML~Cyg) and three Miras (R~Leo, S~CrB, U~Her).  Unlike the
non-vibrational water lines at 22~GHz, 183~GHz (\cite{Cernicharo90}),
321~GHz (\cite{Menten90a}), 325~GHz (\cite{Menten90b}), 439 and
470~GHz (\cite{Melnick93}), the 658~GHz line is not detected as a
maser in star-forming regions.  However, the 658~GHz line is detected
as a thermal line in the CSO Orion~KL 600-720~GHz line survey
(\cite{Schilke01}).  In this paper, we describe the first $1''-2''$
resolution observations of the 658~GHz maser and present new
detections from six additional Mira variables obtained by the
Submillimeter Array (SMA)\footnote{Located on Mauna Kea, Hawaii, the
Submillimeter Array (SMA) is an eight-element interferometer built and
operated as a collaborative project between the Smithsonian
Astrophysical Observatory and the Academia Sinica Institute of
Astronomy \& Astrophysics of Taiwan.}.

\section{Observations}

Summarized in Table~\ref{obs}, the SMA observations were performed
primarily during commissioning periods of the 600 GHz receivers and
their associated IF path (\cite{Hunter02}).  The SMA receivers are
double sideband (DSB), fixed-tuned SIS mixers with an IF output band of
4-6~GHz.  Further description of the SMA antennas and receivers can be
found elsewhere (\cite{Blundell04}).  Prior to January 2005, SMA
observations could only be obtained in one receiver band at a time.
As of January 2005, simultaneous observations of the same target
position with two receiver bands could be obtained: (200~GHz with
600~GHz) or (300~GHz with 600~GHz).  During the observations, in most
cases the 200~GHz band was tuned to the SiO $v$=1, $J$=5-4 maser line
at 215~GHz in lower sideband (LSB). The 658~line was typically tuned
in upper sideband (USB).  The H$_2$O maser line was observed with a
correlator channel spacing of 0.8125 MHz (0.4 km/s).  The compact
configuration was used, with baseline lengths ranging from 16 to 69
meters. Callisto was the flux calibrator.

\begin{table}[htb]
\begin{center}
\caption{Log of significant SMA observations in the 658 GHz water maser line \label{obs}}
\begin{tabular}{|c|c|c|c|l|l|}   
\hline
Date & Log \# & Antennas & $\tau_{\rm 225GHz,zenith}$ & New detections & Other detections\footnotemark[2]\\  
\hline
12 Dec 2002 &  4327 & 3 & 0.03 - 0.05 & U Ori & \\
08 Apr 2004 &  7082 & 4 & 0.07        & R Cas, R Aql & VY CMa, U Her, R Leo\\
          &       &   &             &  & \& W Hya\\
24 Jan 2005 &  8690 & 5 & 0.06        &  & W Hya\\ 
28 Jan 2005 &  8726 & 5 & 0.06        &  & R Leo, VY CMa\\ 
16 Feb 2005 &  8847 & 6 & 0.03 - 0.06 &  & VY CMa\\
24 Aug 2005 &  9962 & 7 & 0.05 - 0.08 &  & VX Sgr\\
14 Dec 2005 & 10802 & 7 & 0.028       & TX Cam, NML Tau &  R Cas\\
11 Jan 2006 & 10721 & 6 & 0.06        & R Hya & VY CMa, R Leo\\ 
\hline
\end{tabular}\\
\footnotemark[2]{Non-detections: IRC+10011, Betelgeuse, IRC+70066, GX Mon, OH231.8+4.2, RS Cnc}
\end{center}
\end{table}


\section{Results}

\subsection{New detections}

\begin{figure}[htb]
\centering
\resizebox{6.5cm}{!}{\rotatebox{0}{\includegraphics{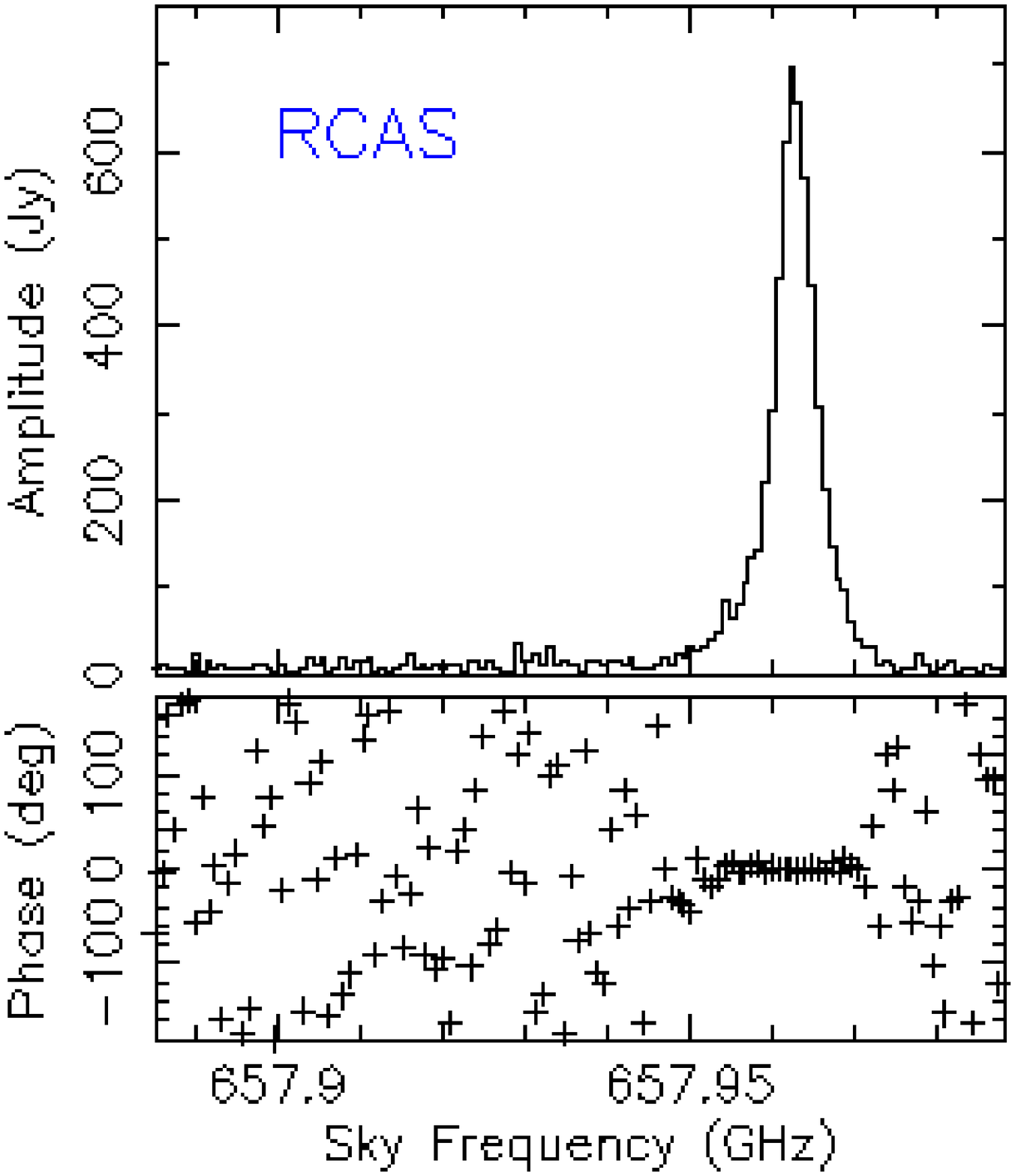}}} 
\resizebox{6.02cm}{!}{\rotatebox{0}{\includegraphics{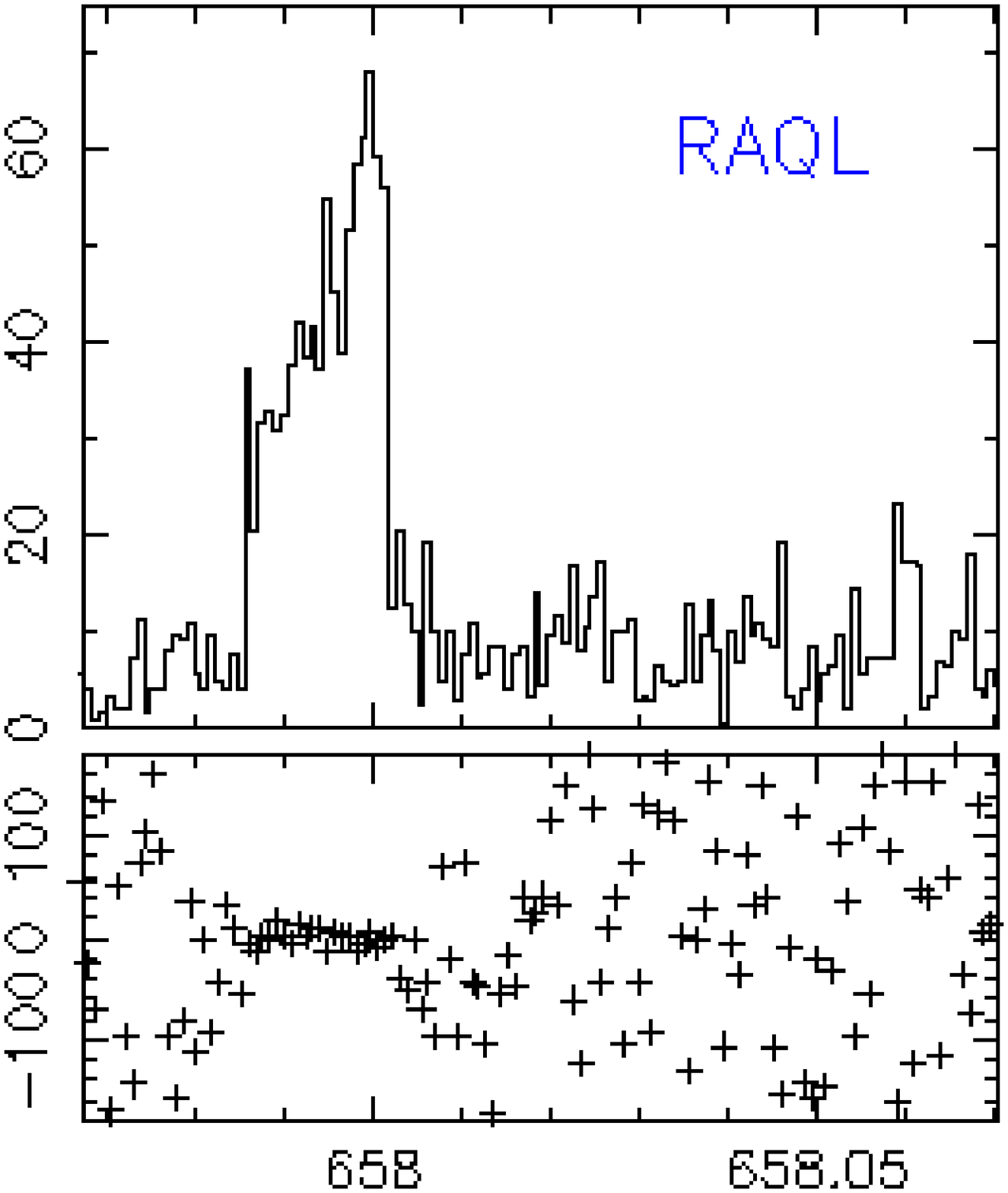}}}  
\caption[]{Calibrated SMA uv spectra of the Mira variables R~Cas (left panel) and R~Aql (right panel) in the 658 GHz H$_2$O maser line. The velocity range shown is 47 km/s (104~MHz).  \label{txcam_nmltau} }
\end{figure}

With the SMA, new detections were obtained of six Mira variables.  As
an example, calibrated uv spectra of R~Cas and R~Aql are shown in
Fig.~\ref{txcam_nmltau}.  

These detections increase the total number
of objects detected in this maser line either by CSO or SMA from 10 to
16.  A complete listing of objects that have been detected in this
line is given in Table~\ref{list}.

\begin{table}[hbt]
\begin{center}
\caption{List of all objects detected in the 658~GHz water maser line (as of March 2007)\label{list}}
\begin{tabular}{|c|l|c|c|c|c|c|}   
\hline
\multicolumn{2}{|c|}{Object} & \multicolumn{2}{|c|}{Position\footnotemark[1]}  & Spectral & Distance & Velocity\\ 
Class & Name & $\alpha$ (J2000) & $\delta$ (J2000) & Type     & (pc) & km/s\\
\hline
                & R Crt    & 11 00 33.85 & $-$18 19 29.6 & M7III, SRb & 186\footnotemark[2] &  8\\ 
semi-regular    & RT Vir   & 13 02 37.98 & +05 11 08.4   & M8III, SRb & 220 & 15\\
pulsating stars & W Hya    & 13 49 02.00 & $-$28 22 03.5 & M7e,  SRa  & 98  & 42\\
                & RX Boo   & 14 24 11.63 & +25 42 13.4   & M7.5, SRb  & 141\footnotemark[2] &  1\\
\hline
            & VY CMa  & 07 22 58.33 & $-$25 46 03.2 & M3/M4II       & 1500\footnotemark[3] & 22\\
hypergiants & VX Sgr  & 18 08 04.05 & $-$22 13 26.6 & M5/M6III, SRc & 1400 & 7\\
            & NML Cyg & 20 46 25.46 & +40 06 59.6 & M6IIIe        & 2000 & $-$1\\
\hline
          & NML Tau & 03 53 28.84 & +11 24 22.6 & M6me & 270   & 32\\
          & TX Cam & 05 00 50.39 & +56 10 52.6 & M8.5 & 380    &  9 \\
          & U Ori & 05 55 49.17 & +20 10 30.7 &  M8III   & 256 & 39 \\
 Mira     & R Leo & 09 47 33.49 & +11 25 43.6 & M8IIIe & 110   & 0\\
variables & R Hya & 13 29 42.78 & $-$23 16 52.8 & M7IIIe & 125 & $-$11  \\
          & S CrB & 15 21 23.96 & +31 22 02.6 & M7e    & 430   & 1 \\
          & U Her & 16 25 47.47 & +18 53 32.9 & M7III  & 347\footnotemark[2]   & $-$15\\
          & R Aql & 19 06 22.25 & +08 13 48.0 &  M7IIIev & 220 &  48 \\
          & R Cas & 23 58 24.87 & +51 23 19.7 & M7IIIe & 184\footnotemark[2] & 26\\
\hline
\end{tabular}\\
\footnotemark[1]{ICRS positions taken from SIMBAD} ~~
\footnotemark[2]{\cite{Colomer00}} ~~
\footnotemark[3]{\cite{Lada78}} 
\end{center}
\end{table}

\subsection{Comparison of H$_2$O and SiO maser line widths} 

\begin{figure}[hbt]
\centering
\resizebox{6.5cm}{!}{\rotatebox{-90}{\includegraphics{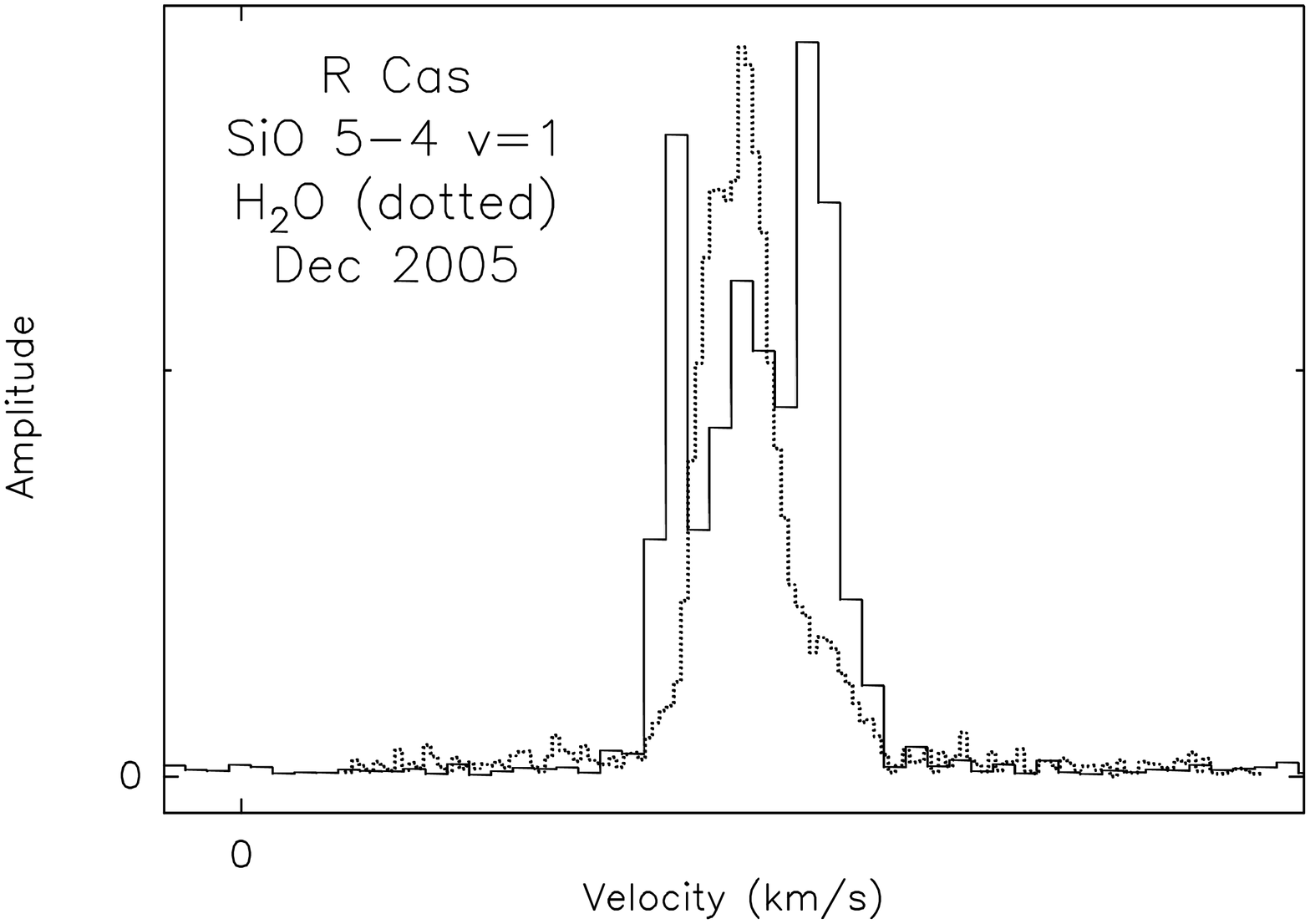}}}
\resizebox{6.5cm}{!}{\rotatebox{-90}{\includegraphics{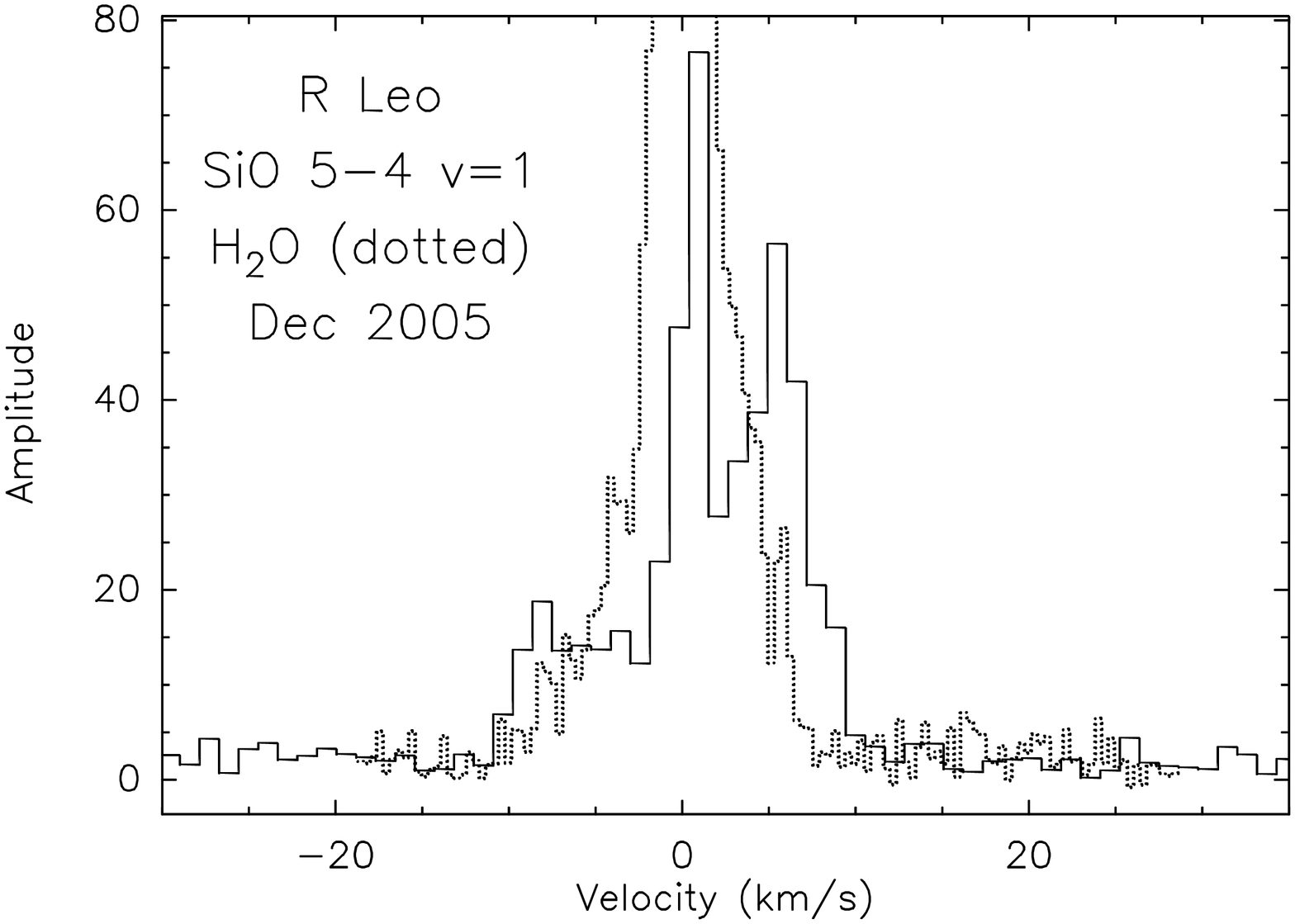}}}\\ 
\resizebox{6.5cm}{!}{\rotatebox{-90}{\includegraphics{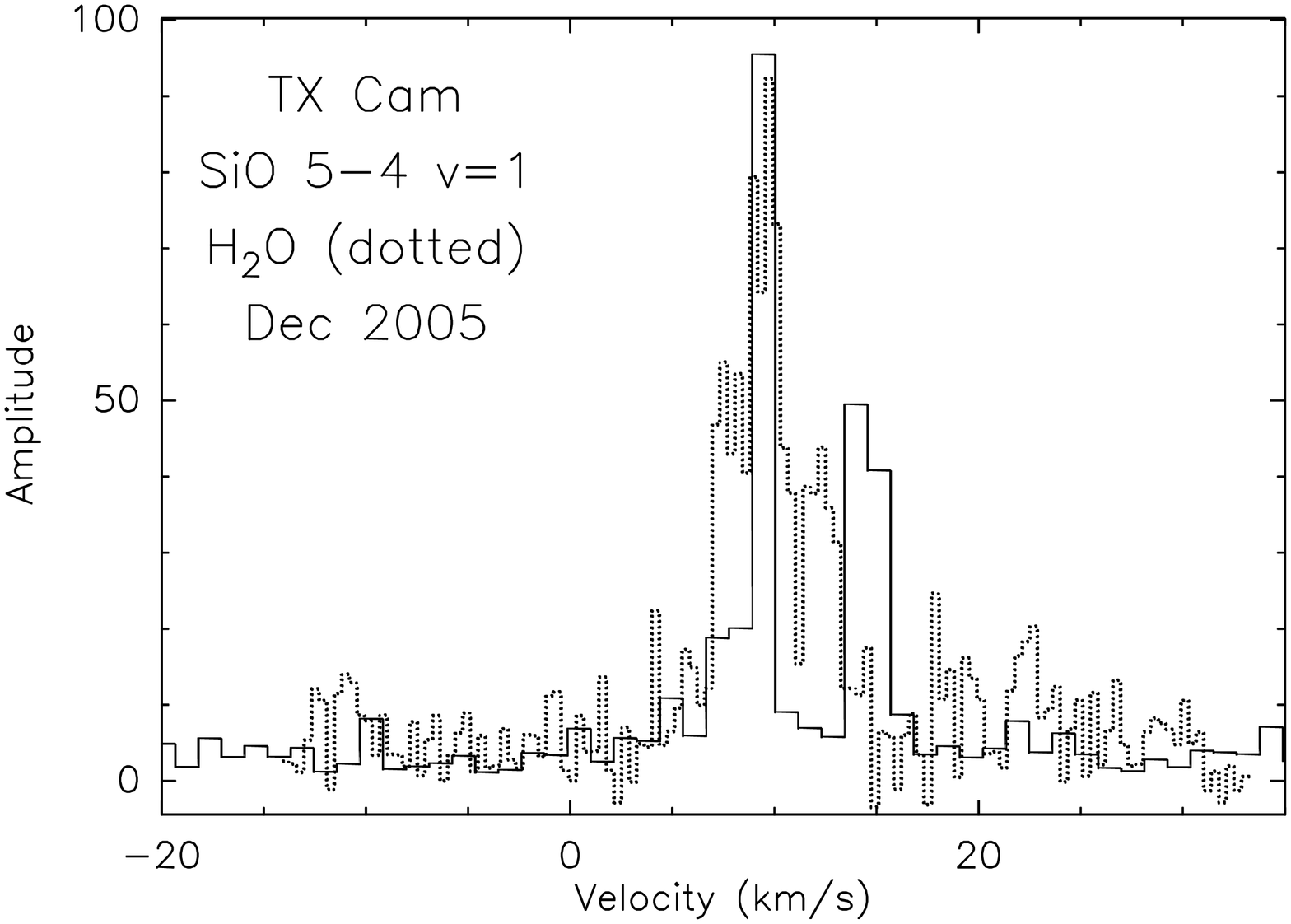}}}
\resizebox{6.5cm}{!}{\rotatebox{-90}{\includegraphics{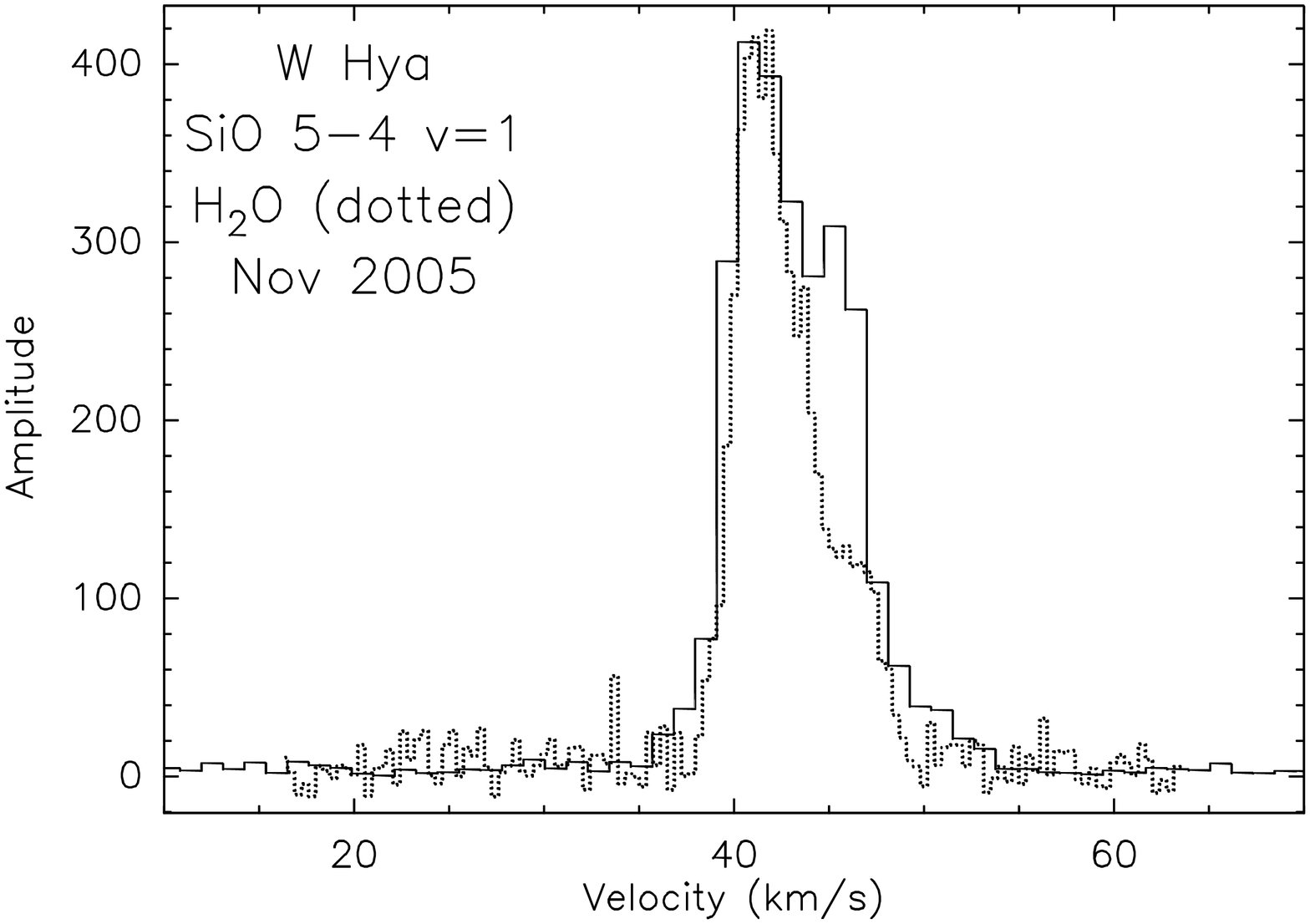}}} 
\caption[]{Simultaneous SMA uv spectra of SiO $J$=5-4, $v=1$ (solid line) and 658 GHz 
H$_2$O (dotted line) in four evolved stars. The channel spacing was 3.25 MHz (4.5 km/s) for SiO, and 0.8125 MHz (0.4 km/s) for H$_2$O. \label{specoverlay}}
\end{figure}

\begin{figure}
\centering
\resizebox{14.0cm}{!}{\rotatebox{-90}{\includegraphics{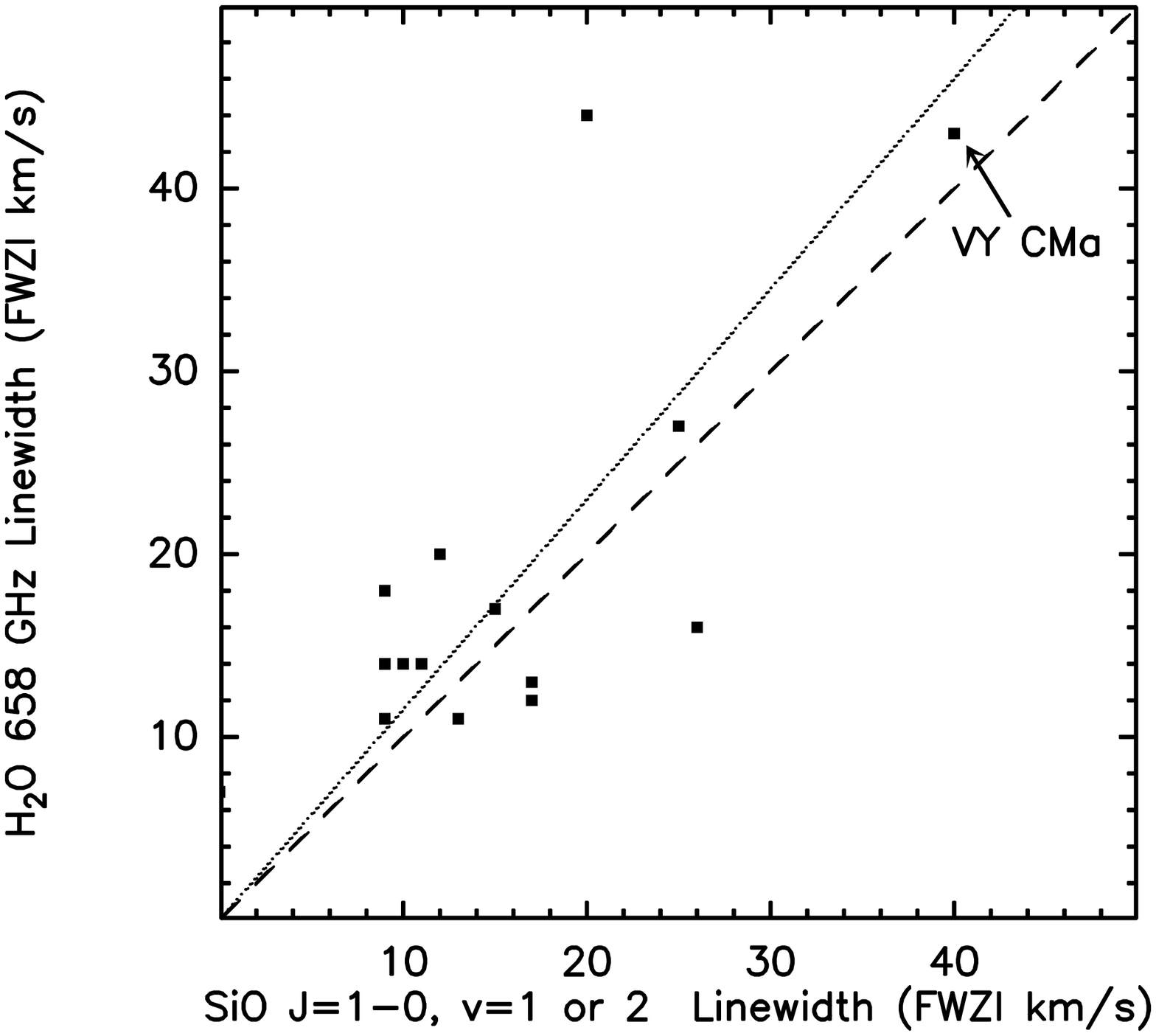}}} 
\caption[]{Scatter plot of FWZI of 658~GHz H$_2$O line vs FWZI of SiO $v=1$, J=1-0 line (or in some cases J=2-1). 
The dashed line has a slope of unity, while the best fit dotted line has a slope of 1.15. \label{scatter}}
\end{figure}

A comparison of the simultaneous spectral observations of the 658~GHz
water maser with the 215~GHz SiO $v=1$ maser in four objects is shown
in Fig.~\ref{specoverlay}.  The emission from these two species arises
from a similar range of velocities, which has been interpreted as
evidence for a common physical origin (\cite{Menten95}), particularly
because the lower level of the SiO $v=1$ maser lies at a similar energy
above the ground state (1792~K) as does the 658~GHz line.  A scatter
plot of the line width (FWZI) of these two masers in all 16 objects is
shown in Fig.~\ref{scatter}.

\subsection{Source size}

\begin{figure}
\centering
\resizebox{12.0cm}{!}{\rotatebox{-90}{\includegraphics{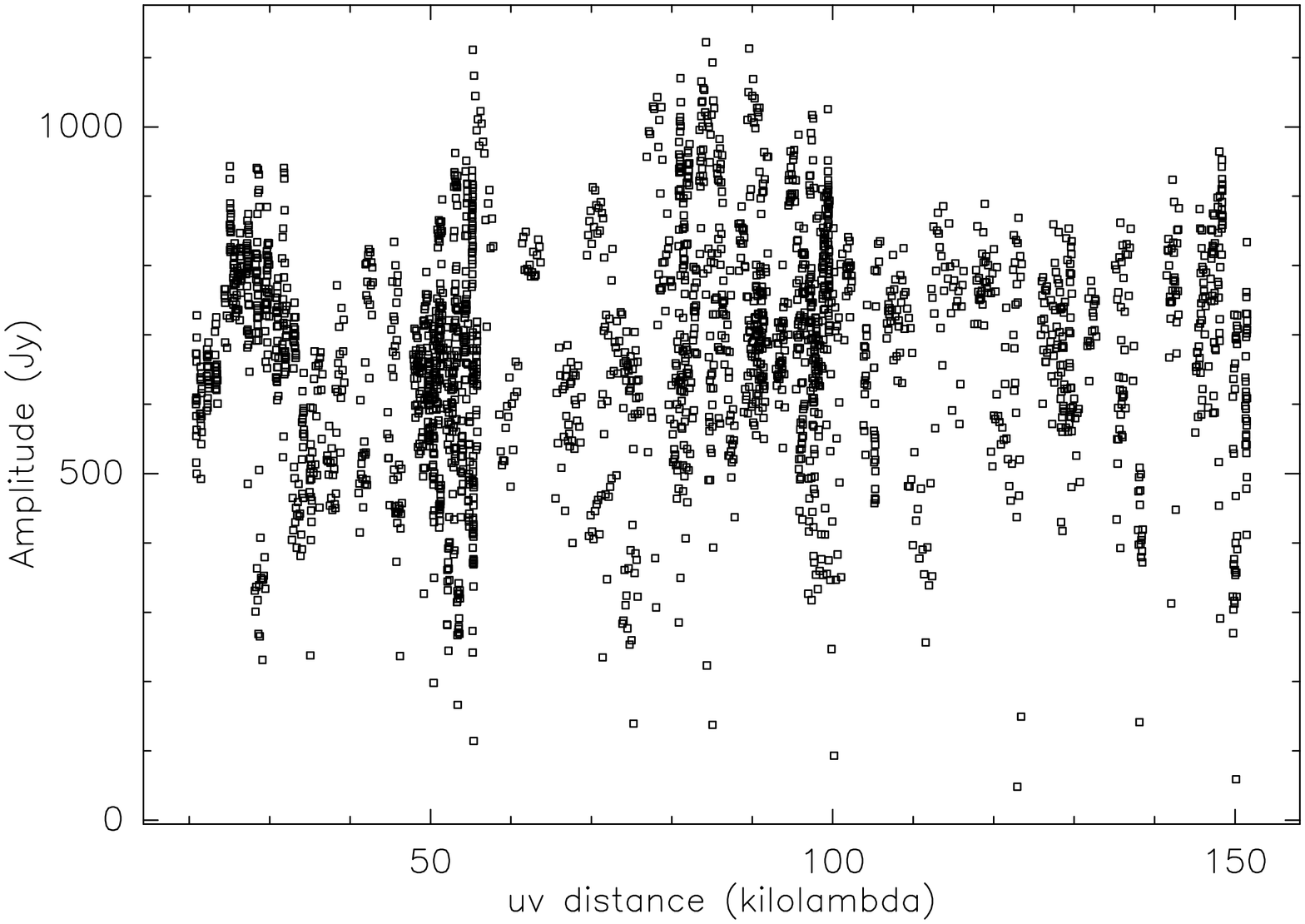}}} 
\caption[]{Calibrated amplitude vs. uv distance for the VY~CMa 658~GHz 
maser observations of 16 Feb 2005 with 15 baselines.\label{uvdist}}
\end{figure}

\begin{figure}
\centering
\resizebox{12.5cm}{!}{\rotatebox{-90}{\includegraphics{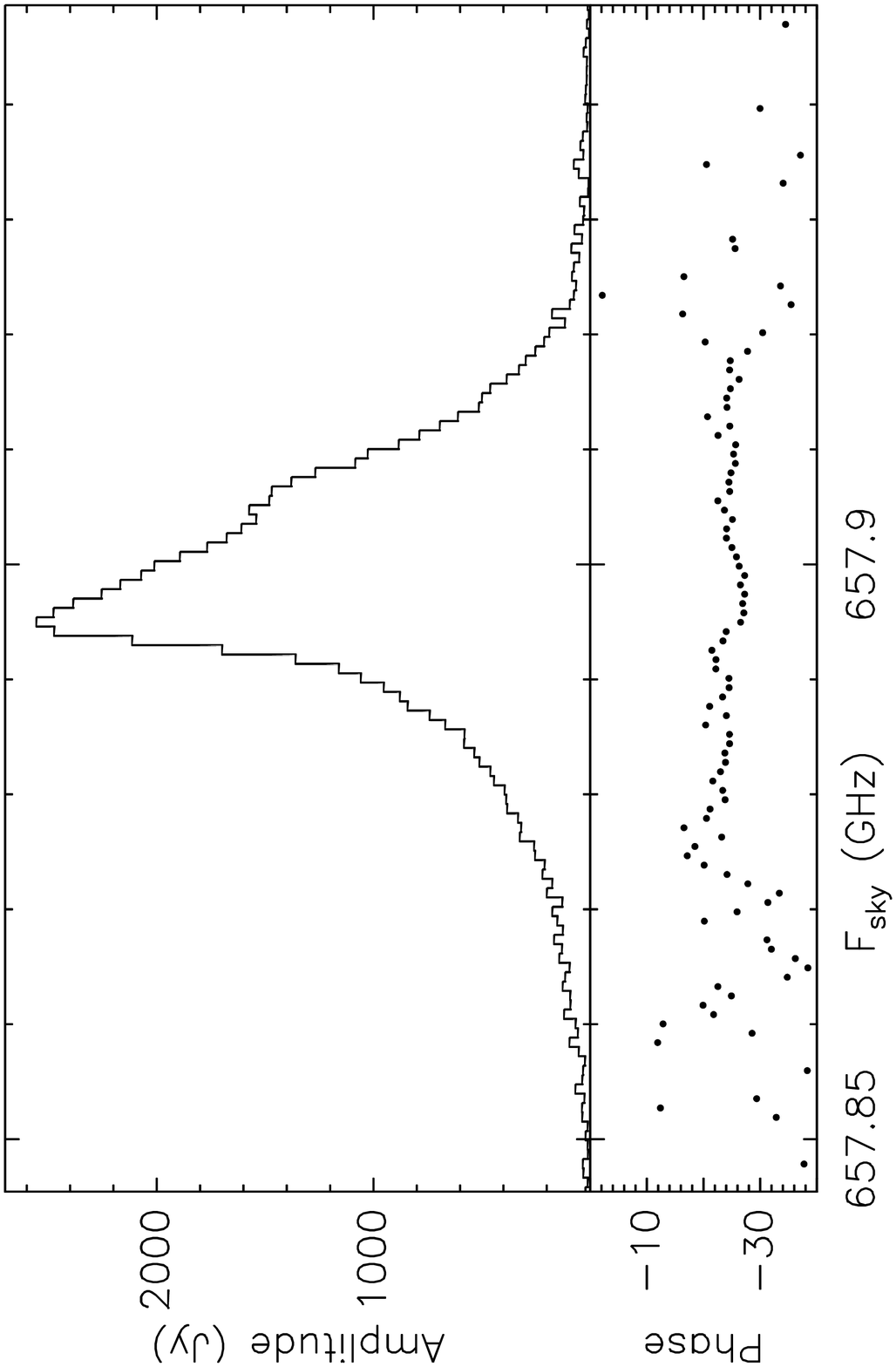}}} 
\caption[]{Calibrated phase in degrees (dots) and amplitude (histogram) vs. frequency for the VY~CMa 658~GHz maser observations of 16 Feb 2005.
This spectrum is a vector average of all 15 baselines.\label{uvspec}}
\end{figure}

To investigate the source size of the 658~GHz masers, we have analyzed
the data for VY~CMa for which one of the highest signal-to-noise
tracks of SMA data exists, amounting to 87 minutes on-source.  The
radio photosphere of this red, luminous ($5\times10^5~L_\odot$ 
e.g. \cite{Humphreys06}) hypergiant
star was found to be unresolved with a $1''$ beam at 22~GHz
(\cite{Lipscy05}).  In the mid-infrared, the same authors measured the
size of the warm dust continuum emission to be $0.3''$ at $17.9~\mu$m.
In the near-infrared, the $K$ band size was measured to be $0.138''
\times 0.205''$ (\cite{Wittkowski98}).  SMA observations at 230~GHz
continuum with a $1.4''$ beam find the emission to be unresolved
(\cite{Muller07}).  Proceeding with the likely assumption that the
658~GHz continuum emission is likewise compact with respect to the
synthesized beam ($1.6'' \times 1.0''$), we generated a continuum
dataset using all the correlator chunks in USB except the one
containing the maser line.  We used that dataset to calibrate the
antenna phases and amplitudes with a point source model, and applied
the solutions to the spectral line data.  A plot of the resulting
amplitude vs. uv distance is shown in Fig.~\ref{uvdist}.  There is no
discernible drop in amplitude with distance, suggesting that the
source remains unresolved at these angular scales.  Further evidence
for this conclusion comes from the uv spectrum (Fig.~\ref{uvspec})
which shows only a very small phase gradient ($<10^\circ$) across the
spectral line.

\section{Phase calibration and phase transfer}

Because they are bright celestial beacons, the 658 GHz water masers
are helpful in debugging problems in the SMA frontends and backends.
The 658~GHz maser line has also served as a successful phase calibrator for
SMA observations of other sources.  \cite{Chen07} used VY~CMa to
calibrate observations of the ultracompact HII region G240.31+0.07,
located less than $5^\circ$ away.  The measured phase of the water
maser emission was used to calibrate the temporal changes in
atmospheric phase plus instrumental phase time while the unresolved
stellar continuum emission ($\sim 9$~Jy) was used to calibrate the
amplitudes.  The 658~GHz maser line has also been used to test the
effectiveness of phase transfer between the 200 and 600~GHz bands on
the SMA.  Using observations of the star W~Hydra on 28 Jan 2005, the
antenna-based phase solutions for the 215~GHz SiO maser emission and
the 658~GHz H$_2$O maser emission were compared over a period of
several hours and shown to change in a ratio close to that expected
theoretically (\cite{Hunter05}).  At SMA, the usage of this technique
is limited due to some remaining instrumental thermal drifts and phase
jumps currently under investigation (\cite{Kubo06}).  However, the
phase transfer technique is expected to be relied upon heavily by ALMA
(\cite{Laing04}), typically from Band 3 (84-116~GHz) where bright
quasars are more numerous, to higher frequency bands where they are
not (\cite{Butler03}).

\section{Future prospects}

\begin{figure}[t]
\centering
\resizebox{13.1cm}{!}{\rotatebox{-90}{\includegraphics{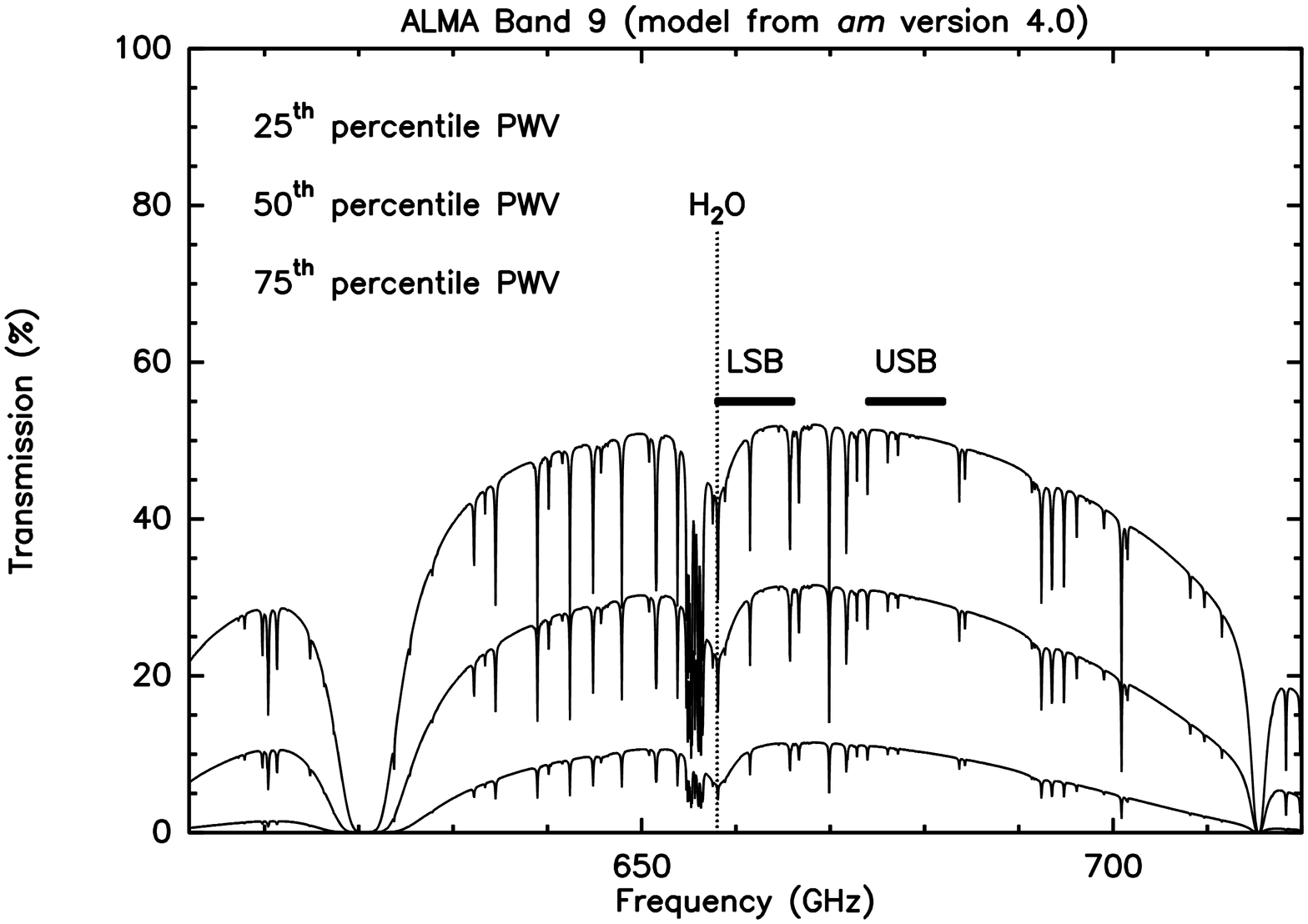}}} 
\caption[]{Proposed tuning for 660~GHz continuum observations with ALMA, 
including the 658~GHz line in the LSB passband in order to use an evolved star
as a phase calibrator near the target source.  The atmospheric model
curves are from the {\it am} package (\cite{Paine06}).  The top curve
is 25th \%ile conditions, the middle curve is 50th \%ile, and the
bottom curve is 75th \%ile.
\label{alma}}
\end{figure}

At the present time, the SMA has the potential to observe the 658~GHz
maser sources with a beamsize as small as $0.15''$, where it may begin
to resolve the emission as a phase gradient in velocity.  For example,
43~GHz VLBI observations of TX~Cam show a ring of SiO masers with a
diameter of $\sim 0.04''$ (\cite{Diamond03}).  After its launch in
2008, the Band 2 receiver (\cite{Teipen05}) of the Heterodyne
Instrument for the Far-Infrared (HIFI) aboard the Herschel satellite
(\cite{Poglitsch06}) will provide the opportunity to perform a
sensitive survey of 658~GHz maser emission in a large number of stars.
Beginning around 2010, ALMA will become the first telescope capable of
observing all five known (sub)millimeter transitions of
vibrationally-excited water (\cite{Wootten07}). This capability will
be important for obtaining near-simultaneous observations of all these
lines in order to accurately determine the temperature and density of
the emitting gas.  As shown in Fig.~\ref{alma}, the Band 9 receivers
of ALMA are DSB and can be tuned to provide the 658~GHz water maser
line in the LSB and a 4~GHz region of continuum in the USB that is
nearly free of atmospheric ozone absorption features.

\end{document}